\documentclass[iop,apj]{emulateapj}
\usepackage{booktabs}

\usepackage{graphicx,epstopdf,hyperref,color}

\newcommand{\Teff}{\mbox{$T_{\mathrm{eff}}$}}

\begin{document}
\title{Machine learning Applied to Star-Galaxy-QSO Classification and Stellar Effective Temperature Regression}

   \author{Yu Bai\altaffilmark{1}}
   \author{JiFeng Liu\altaffilmark{1,2}}
   \author{Song Wang\altaffilmark{1}}
   \author{Fan Yang\altaffilmark{2}}

\altaffiltext{1}{Key Laboratory of Optical Astronomy, National Astronomical Observatories, Chinese Academy of Sciences,
       20A Datun Road, Chaoyang Distict, Beijing 100012, China; ybai@nao.cas.cn}
\altaffiltext{2}{College of Astronomy and Space Sciences, University of Chinese Academy of Sciences, Beijing 100049, China}

\begin{abstract}
In modern astrophysics, the machine learning has increasingly gained more popularity with its incredibly powerful ability
to make predictions or calculated suggestions for large amounts of data.
We describe an application of the supervised machine-learning algorithm, random forests (RF), to the star/galaxy/QSO
classification and the stellar effective temperature regression based on the combination of LAMOST and SDSS spectroscopic data.
This combination enable us to obtain reliable predictions with one of the largest training sample ever used.
The training samples are built with nine-color data set of about three million objects for the classification and
seven-color data set of over one million stars for the regression. The performance of the classification and regression
is examined with the validation and the blind tests on the objects in the RAVE, 6dFGS, UVQS and APOGEE surveys.
We demonstrate that the RF is an effective algorithm with the classification accuracies higher than 99\% for the stars and the galaxies, and
higher than 94\% for the QSOs. These accuracies are higher than the machine-learning results in the former studies.
The total standard deviations of the regression are smaller than 200 K that is similar to those of some spectrum-based methods.
The machine-learning algorithm with the broad-band photometry provides us a more efficient approach to deal with massive amounts of
astrophysical data than traditional color-cuts and SED fit.
\end{abstract}


\section{Introduction}
\label{sect:intro}
Nowadays, astronomy and cosmology is concerned with the study and characterization of millions of objects,
which could be quickly identified with their optical spectra. However, billions of sources in wide-field
photometric surveys cannot be followed-up spectroscopically, and an appropriate identification of various
source types is complicated \citep{Krakowski16}. In a traditional way, a separator between stars and galaxies is a
morphological measurement \citep{Vasconcellos11}, but we quickly reach the limit due to low image resolution.
Another separation involves magnitudes and colors criteria, but the criteria become too
complex to be described with functions in a multidimensional parameter space.

However, this parameter space can be effectively explored with machine-learning algorithms, e.g., the
support vector machines (SVM; \citealt{Cortes95,Kovacs15,Krakowski16}), RF \citep{Breiman01,Yi14,Reis18} and $k$-nearest neighbours \citep{Fix51,Garcia18}.
Machine learning teaches computers to learn from
"experience" without relying on a predetermined equation or an explicit program. It finds natural patterns in
data that generate insight and help us make better decisions and predictions \footnote{https://www.mathworks.com/solutions/machine-learning.html}.
Machine-learning algorithms have helped us to deal with complex problems in astrophysics,
e.g., automatic galaxy classification \citep{Huertas08,Huertas09}, the Morgan-Keenan spectral classification (MK; \citealt{Manteiga09,Navarro12,Yi14}),
variable star classification \citep{Pashchenko18} and spectral feature recognition for QSOs \citep{Parks18}.

The "experience" used for the machine learning is also known as training data, which is the key to make effective predictions.
The classification from spectroscopic surveys is an ideal training data due to its high reliability.
Several works have been done that explore the performance of the star/galaxy/QSO classification
(e.g., \citealt{Suchkov05,Ball06,Vasconcellos11,Kovacs15,Krakowski16}).
In these studies, the machine-learning classifiers were built with photometric colors and spectroscopic classes, and shown more
accurate prediction than other traditional methods such as color cuts \citep{Weir95}.
However, there are still some locations in the multi-color space that weren't explored by the classifiers,
owing to the small size of spectroscopic sample. Therefore, a machine-learning classifier built from a large
spectroscopic sample is required to cover a more complete multi-color space, and further to yield accurate classification
for billions of sources.

After separating stars from galaxies and QSOs, we want to understand their nature.
The stellar spectral classification, the MK spectral types, is the fundamental reference frame of stars.
However, the method for the MK classification is based on features extracted from the spectra
\citep{Manteiga09,Daniel11,Navarro12,Garcia18}, which limits the application to the stars with high
signal-to-noise ratio.
On the other hand, the spectral features of different types could be very similar, and thus it is difficult
to make clear cuts for different spectral types \citep{Liu15}.  An alternative method is estimating the
effective temperature with multi colors, which only requires photometric data and has the ability to cover a
greater area of the sky.
Some theoretical studies have indicated that combining broad-band photometry allows atmospheric parameters
and interstellar extinction to be determined with fair accuracy \citep{Bailer13,Allende16}. However,
there is still no research to test its validation with real observational data.

In this paper, we take advantage of the archive data from the SDSS and the LAMOST surveys (Seciton \ref{data})
to build the star/galaxy/QSO classifier (Section \ref{class}) and stellar effective temperature regression
(Section \ref{regress}) based on one of the largest machine-learning sample.
The validation and the blind tests are applied to explore the performance of the prediction in Section \ref{class} and \ref{regress}.
In Section \ref{dis}, we present the comparisons with other machine-learning methods, and
application of the SED fit to the real observational data. A summary and future work are given in Section \ref{sum}.

\section{Data}\label{data}
\subsection{SDSS and LAMOST Spectroscopic Surveys}
\label{sect:Ssur}
Sloan Digital Sky Survey (SDSS) is an international project that has made the most detailed 3D maps of our Universe.
The fourth stage of the project (SDSS-IV) started in 2014  July, with plans to continue until mid-2020 \citep{Blanton17}.
The automated spectral classification of the SDSS-IV is determined with
chi-square ($\chi^2$) minimization method, in which the templates are constructed by performing a rest-frame principal-component
analysis \citep{Bolton12,Blanton17}. The first data release in the SDSS-IV, DR13, includes over 4.4 million sources,
in which galaxies comprise 59\%, QSOs 23\%, and stars 18\% \citep{Albareti17}.

Another on-going spectroscopic survey is undertaken by the Large Sky Area Multi-Object Fiber Spectroscopic Telescope
(LAMOST, \citealt{Cui12,Zhao12}), which mainly aims at understanding the structure of the Milky Way \citep{Deng12}.
In 2016.06.02, the LAMOST finished the forth year survey (the forth data release; DR4), and
has obtained spectra of more than 7.6 million sources included 91.6\% stars, 1.1\% galaxies, and 0.2\% QSOs.
The LAMOST 1D pipeline recognizes the spectral classes by applying a cross-correlation with templates \citep{Luo15}.
An additional independent pipeline and visual inspection are carried out in order to double check the galaxies
and QSOs identification. We here adopt SDSS DR13 plus LAMOST DR4, since they are matched on the time of the data releasing.

\begin{figure}
   \centering
   \includegraphics[width=0.45\textwidth]{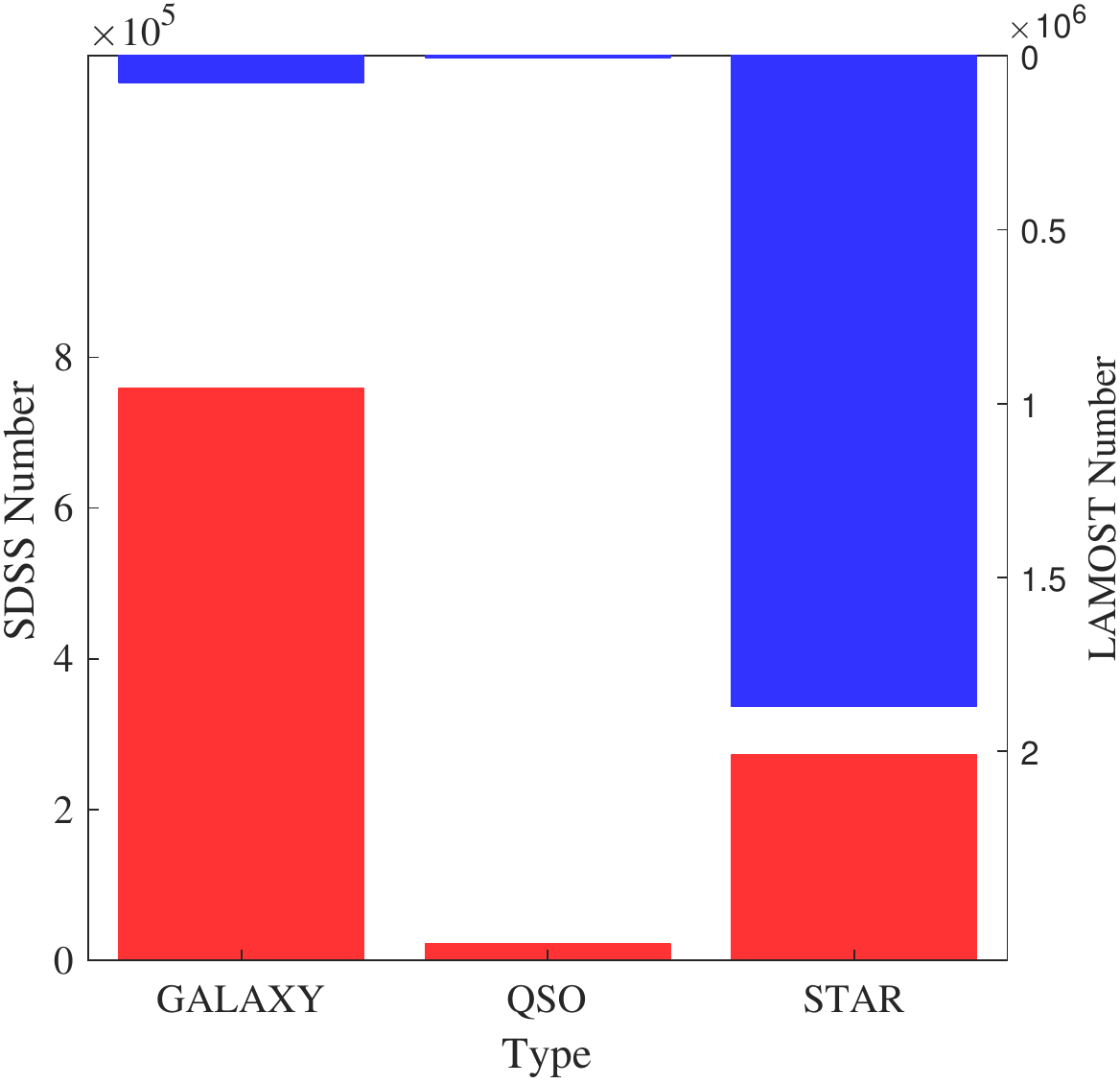}
   \caption{The comparison of the object types between SDSS DR13 (red) and LAMOST DR4 (blue) spectroscopic survey.
   \label{Typ}}
\end{figure}

Figure \ref{Typ} shows the comparison between the two spectroscopic surveys. The objects of LAMOST is dominated by
stars, while over half of the objects in SDSS are galaxies. The combination of the two surveys can provide a more balanced
and larger training sample for the classification.
In order to add more QSO samples, we adopt the 13th edition of quasars and active nuclei catalogs (Veron13, \citealt{Veron10}),
which includes 23,108 samples.
The priority of the catalogs is Veron13, SDSS and LAMOST, if some objects are included in more than one catalog.

\subsection{SDSS and $WISE$ Photometric Surveys}
The combination of optical and infrared (IR) data on huge numbers has been proved to be valid in the star/galaxy
classification \citep{Baldry10,Henrion11} and stellar parameter determination \citep{Allende16}.
The SDSS has imaged over 31 thousand square degrees in five broad bands ($ugriz$). The DR13 includes photometry
for over one billion objects. The Wide-field Infrared Survey Explorer ($WISE$; \citealt{Wright10}),
performed an all-sky survey with images in 3.4, 4.6, 12 and 22 $\mu$m and yielded more than half a billion objects.

In order to obtain the training sample, we extract the objects with the available model magnitudes in $g$, $r$, $i$
bands for the SDSS and LAMOST spectroscopic surveys, and cross identify them with the $WISE$ All-Sky Data Release catalog
with the help of the Catalog Archive Server Jobs (CasJobs)\footnote{http://skyserver.sdss.org/CasJobs/}. Similar to
\citep{Krakowski16}, we use w1(2)mpro magnitudes.
The $J$, $H$, and $K$ magnitudes are also extracted in order to cover the near IR bands.
Our selection required the object with zWarning = 0 for the SDSS objects, and S/N ratios higher than 2 in the $W$1 and $W$2 bands.
We adopt the \textrm{w?mag13} as the indicators for the extended objects \citep{Bilicki14,Krakowski16,Kurcz16,Solarz17}, which is defined as
\begin{equation}
\mathrm{w?mag13} = \mathrm{w?mag\_1} - \mathrm{w?mag\_3},
\end{equation}
where \textrm{w$?$mag\_1} and \textrm{w$?$mag\_3} are magnitudes measured with circular apertures of radii of {5.5\arcsec} and {11\arcsec}.
The question mark is the channel number in the catalog.

\section{Star/Galaxy/QSO Classification}\label{class}
Classification lies at the foundation of astronomy, and it is the beginning of understanding the relationships between
disparate groups of objects and identifying the truly peculiar ones \citep{Gray09}. In this section, we present the machine-learning
method and the performance tests of our classifier.
\subsection{Method}
\begin{figure}
   \centering
   \includegraphics[width=0.4\textwidth]{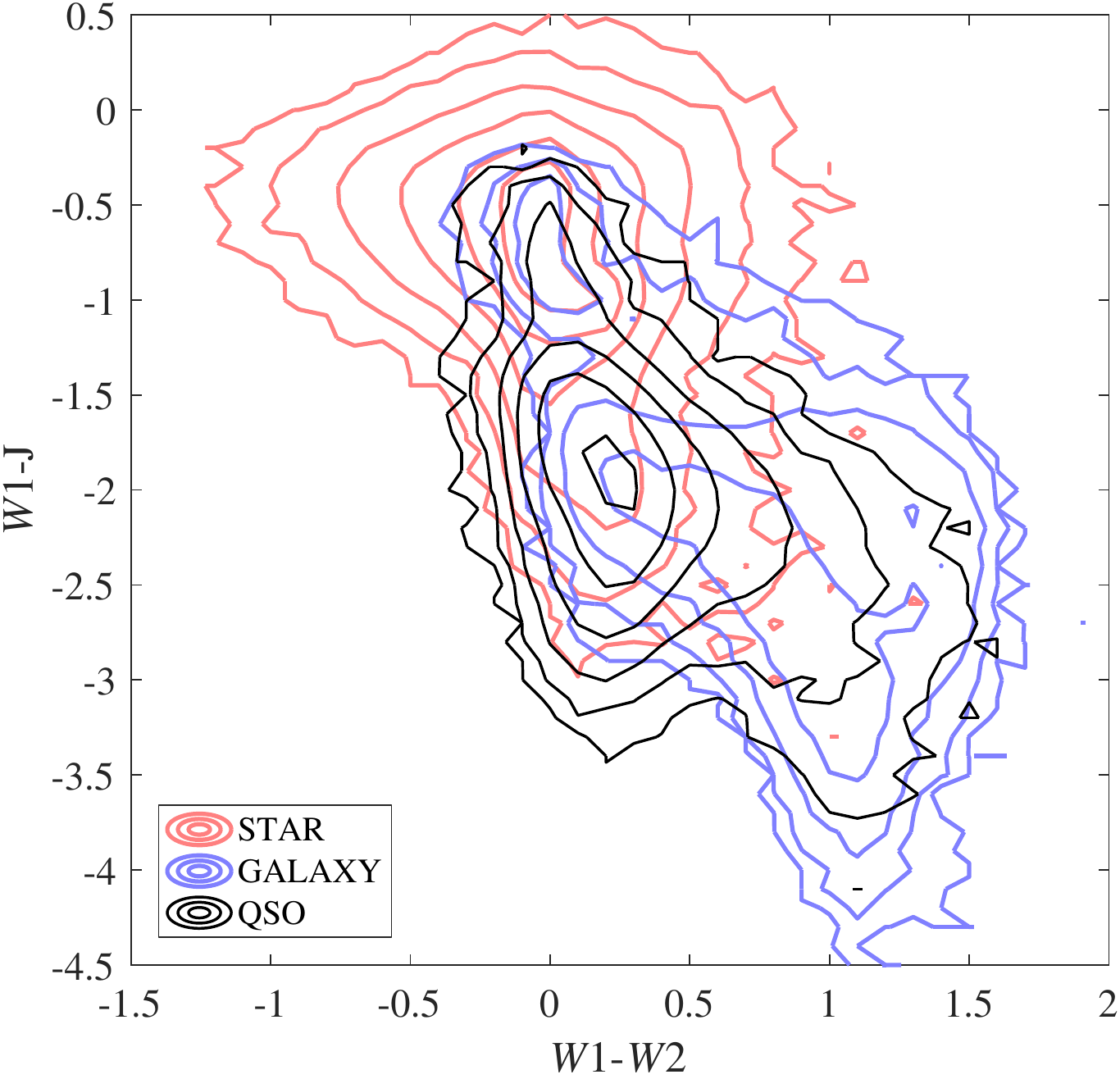}
   \caption{The color-color diagram for stars (red), galaxies (blue) and QSOs (black) in our training sample.
   \label{W12J}}
\end{figure}

We use the CasJobs to cross identify the photometric data with the spectral catalogs of LAMOST, SDSS and Veron13.
The result has 2,973,855 objects, included 2,123,831 stars, 806,139 galaxies and 43,885 QSOs.
We present the color-color diagram in Figure \ref{W12J} that is often used for the star-galaxy
separation (e.g., \citealt{Jarrett11,Goto12,Ferraro15}).
The contours of the three classes overlap in the color-color diagram.
Neither the cut $W$1$-W$2 = 0.8 \citep{Stern12,Yan13},
nor $W$1$-J$ = $-$1.7 \citep{Goto12} could provide a clear cut to classify
the stars, galaxies and QSOs.

We build a nine-dimensional color space, $g-r$, $r-i$, $i-J$, $J-H$, $H-K$, $K-W$1, $W$1$-$$W$2,
w1mag13, and w2mag13.
Each object is weighted with the quadratic sum of their photometric uncertainty.
The holdout validation is applied to test the total accuracies of different machine-learning algorithms,
in which a random partition of 20\% is held out for the prediction and the rest is used to train the classifier.

Table \ref{Table0} lists the accuracies and time costs of different algorithms for the 20\% held out samples.
Since the validation is applied, the time costs are approximate. 
The RF algorithm \citep{Breiman01} shows the best performance on the time cost (57 min
\footnote{CPU: i7-3770 @ 3.40GHz, 5 workers for the parallel computing.}) and the total accuracy (99.2\%).
Other methods, for example the $k$-nearest neighbor and the SVM, either cost more time to
build the classifiers or show lower total accuracies.

\begin{deluxetable}{lcr}
\tablecaption{The accuracies and time costs of different algorithms \label{Table0}}
\tablehead{Algorithm & Accuracy [\%] & Time cost$^1$
           }
\startdata
Simple Tree$^2$           & 97.6 & minutes\\
Medium Tree$^3$           & 98.6 & minutes\\
Complex Tree$^4$          & 98.8 & minutes\\
Linear Discriminant$^5$   & 98.3 & a minute\\
Quadratic Discriminant$^6$& 98.2 & a minute\\
Fine KNN$^7$              & 98.7 & a hour\\
Medium KNN$^8$            & 99.1 & a hour\\
Coarse KNN$^9$            & 99.0 & hours\\
Cosine KNN$^{10}$            & 99.0 & hours\\
Cubic KNN$^{11}$          & 99.1 & hours\\
Weighted KNN$^{12}$       & 99.0 & hours\\
RF                        & 99.2 & hours\\
Linear SVM$^{13}$         & 98.9 & a week\\
Quadratic SVM$^{14}$      & 90.6 & a week\\
Fine Gaussian SVM$^{14}$  & 98.9 & a week\\
Cubic SVM$^{14}$          & 72.9 & a week\\
Medium Gaussian SVM$^{14}$& 99.1 & a week\\
Coarse Gaussian SVM$^{14}$& 99.2 & a week\\
\enddata
\tablecomments{1. One worker for the parallel computing.
               2. Few leaves to make coarse distinctions between classes.
               3. Medium number of leaves for finer distinctions between classes.
               4. Many leaves to make many fine distinctions between classes.
               5. Creates linear boundaries between classes.
               6. Creates nonlinear boundaries between classes.
               7. Finely detailed distinctions between classes.
               8. Medium distinctions between classes.
               9. Coarse distinctions between classes.
               10. Medium distinctions between classes, using a Cosine distance metric.
               11. Medium distinctions between classes, using a cubic distance metric.
               12. Medium distinctions between classes, using a distance weight.
               13. Makes a simple linear separation between classes.
               14. Makes a nonlinear separation between classes.
}
\end{deluxetable}

The working theory of the RF is that it builds an ensemble of unpruned decision trees
and merges them together to get a more accurate and stable prediction.
The algorithm consists of many decision trees and outputs the class
that is the mode of the class output by individual trees \citep{Breiman01,Gao09}.
The RF are often used when we have very large training datasets and a very large
number of input variables.
One big advantage of RF is fast learning from a very large number of data.
\citet{Gao09} listed many other advantages of the RF.

\begin{figure}
   \centering
   \includegraphics[width=0.4\textwidth]{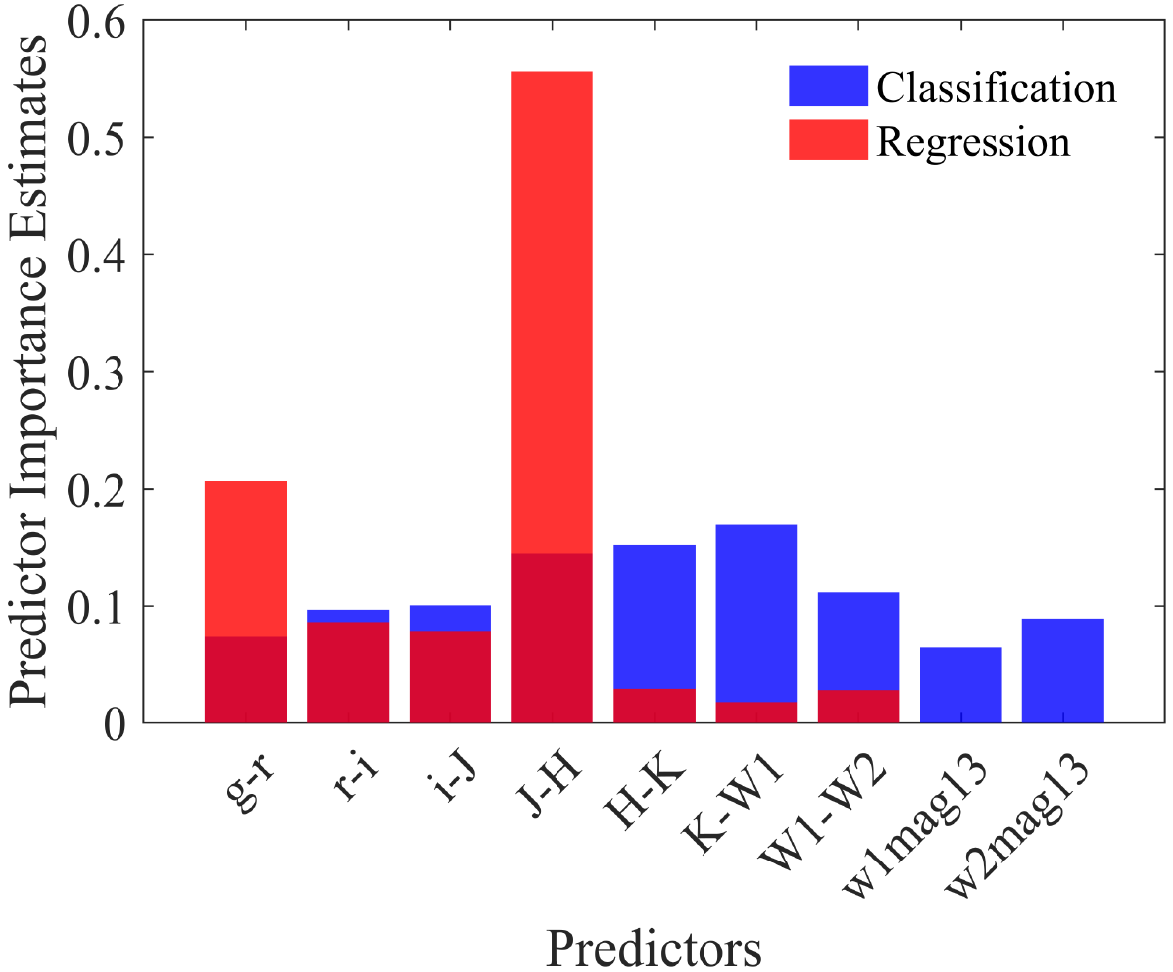}
   \caption{The predictor importance estimates for the classification and the regression.
   \label{Imp}}
\end{figure}

\begin{figure}
   \centering
   \includegraphics[width=0.23\textwidth]{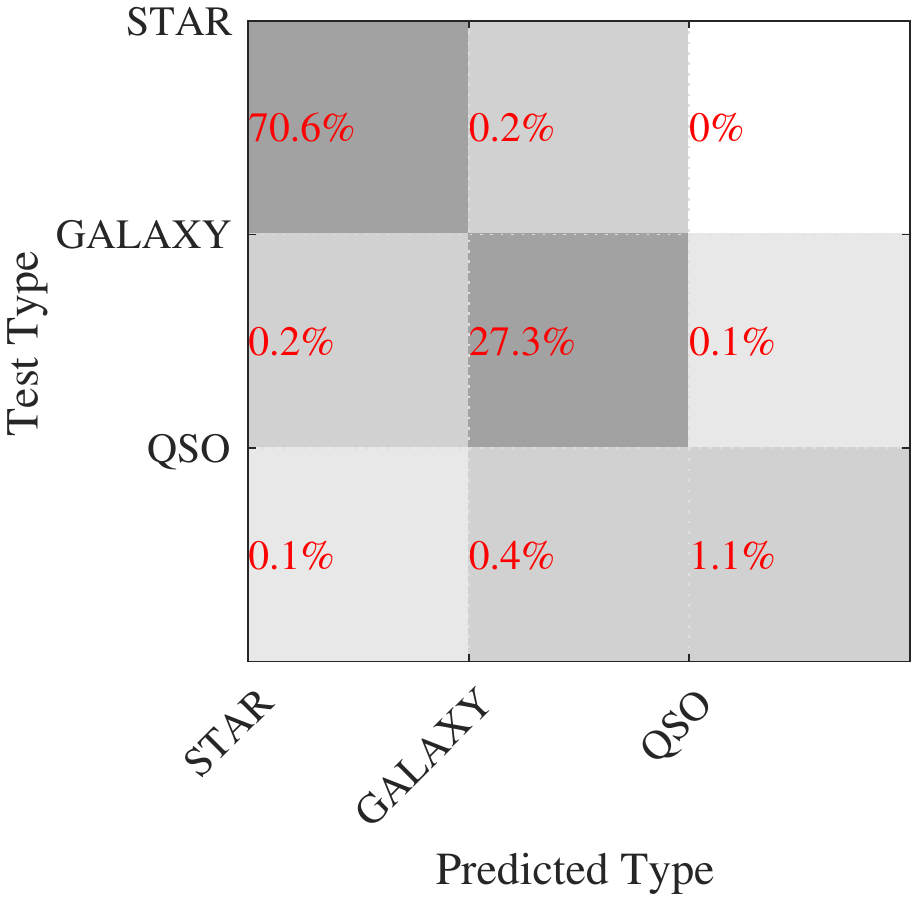}
   \includegraphics[width=0.23\textwidth]{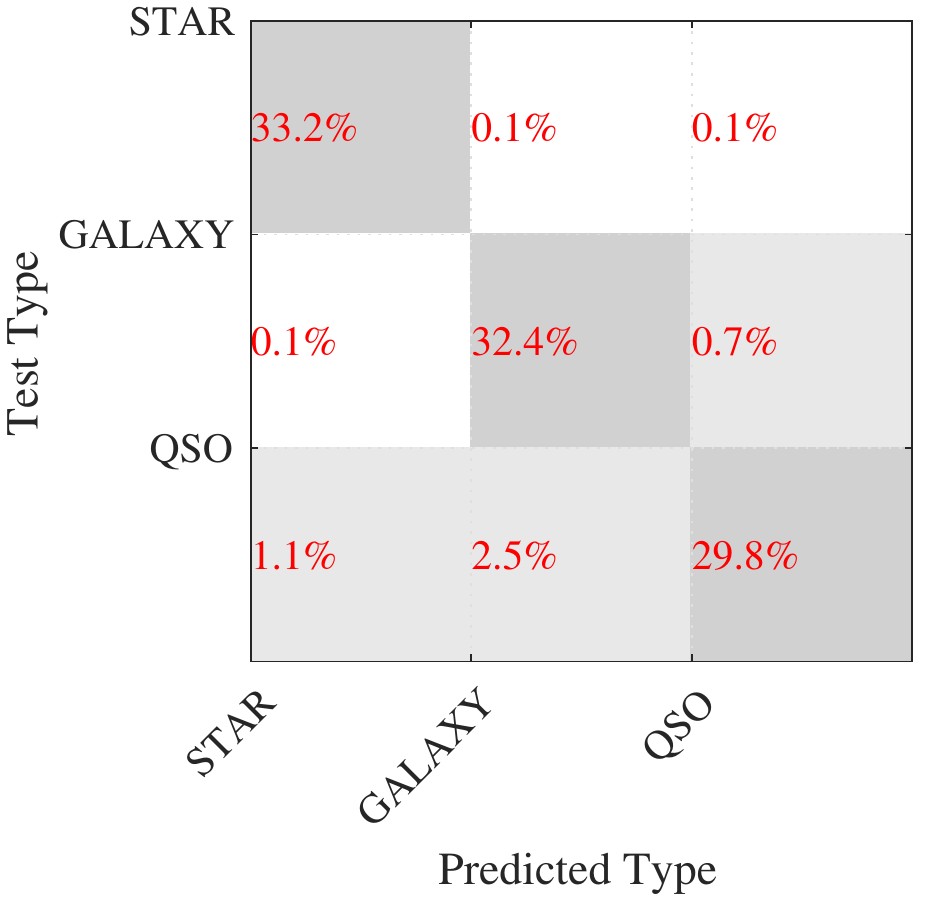}
   \caption{The comparison of the confusion matrixes for two classifiers: one using all samples and the other using
   samples with the same numbers of the different classes.
   \label{cfm}}
\end{figure}

After selecting the best algorithm, we apply the holdout validation and test the RF classifier ten times.
The average accuracy is 99\% and the result is shown in Table~\ref{Table1} and Figure \ref{cfm}.
In the classifier, the contributions of the nine colors are different, which could be described by the predictor
importance estimates (Figure \ref{Imp}).
We find that the IR colors play an important role in our classifier, which is similar to the result of \citet{Krakowski16}.

We adopt the measures defined by \citet{Soumagnac15} to show the performance of the classifier. These measures have been
used in other machine-learning studies \citep{Kovacs15,Krakowski16}: completeness ($c$), and purity ($p$) for star, galaxy
and QSO samples. We use the following equations (here for galaxies):
\begin{equation}
c_{g} = \frac{\textrm{TG}}{\textrm{TG+FGS+FGQ}},
\end{equation}
\begin{equation}
p_{g} = \frac{\textrm{TG}}{\textrm{TG+FSG+FQG}}.
\end{equation}
The FGS and FGQ are the numbers of galaxies misclassified as stars and QSOs, and FSG and FQG are the numbers of stars and QSOs
misclassified as galaxies.
The QSO sample shows the lowest completeness and purity, probably due to its smallest sample size.

In order to test this effect, we normalize the sample sizes of the three classes,
and apply the 20\% holdout validation again (about 8,800 objects in each type for the testing).
The average result is shown in the Table \ref{Table1} and the right panel of Figure \ref{cfm}.
The three classes have similar percentages of the completeness and purity.
This result implies that we could not judge the performance of the classifier only by these measures.

We also apply the magnitude binnings suggested by \citep{Krakowski16} to test the completeness,
since different magnitudes stand for different stellar and galactic types and distances.
The binnings are 12 $< W$1 $<$ 13, 13 $< W$1 $<$ 14, 14 $< W$1 $<$ 15, and 15 $< W$1 $<$ 16.
The completenesses for stars, galaxies and QSOs are similar to those calculated without binning.
On the even samples, the low performance of the galaxy sample is probably due to the relatively high
contamination from the QSO sample rather than the lost information of the galaxy sample.

\begin{deluxetable}{c|cc|cc}
\tablecaption{The comparison of the average performance \label{Table1}}
\tablehead{& \multicolumn{2}{|c|}{All Samples} & \multicolumn{2}{|c}{Uniform Samples}  \\
           & $c$ [\%] & $p$ [\%] & $c$ [\%] & $p$ [\%]
           }
\startdata
Stars      & 99.6     & 99.7     & 99.6     & 96.5 \\
Galaxies   & 98.9     & 97.8     & 97.6     & 92.5 \\
QSOs       & 71.9     & 88.5     & 88.9     & 97.4 \\
\hline
Accuracy         & \multicolumn{2}{|c|}{99}  & \multicolumn{2}{|c}{95}
\enddata
\tablecomments{All samples: the classifier using all samples. Uniform samples: the classifier using samples with the same numbers of the different classes.
              $c$ = completeness, and $p$ = purity.}
\end{deluxetable}

\subsection{Blind Test}
This section describes various tests using the classifier made from the LAMOST, SDSS and Veron13. These tests
allow us to quantify the performance of the classification.

\subsubsection{6dF Galaxy Survey}
The 6dF Galaxy Survey has (6dFGS) mapped the nearby universe over nearly half the sky \citep{Jones04,Jones09}.
The final redshift release of the 6dFGS contains 124,647 spectrally identified galaxies. We match the galaxies
with the SDSS and $WISE$ archive data, which yields 12,300 galaxies. We then remove
the galaxies which are used to build the classifier, and there are 8,382 galaxies left. We feed the classifier with nine colors of these
galaxies, and obtain the predicted types. The classifier can output three scores for each entry, corresponding
to the possibilities for star, the galaxy and the QSO. The type with the largest score is adopted as the predicted type. The
classifier also output the standard deviation ($\sigma$) for each score.

\begin{figure}
   \centering
   \includegraphics[width=0.45\textwidth]{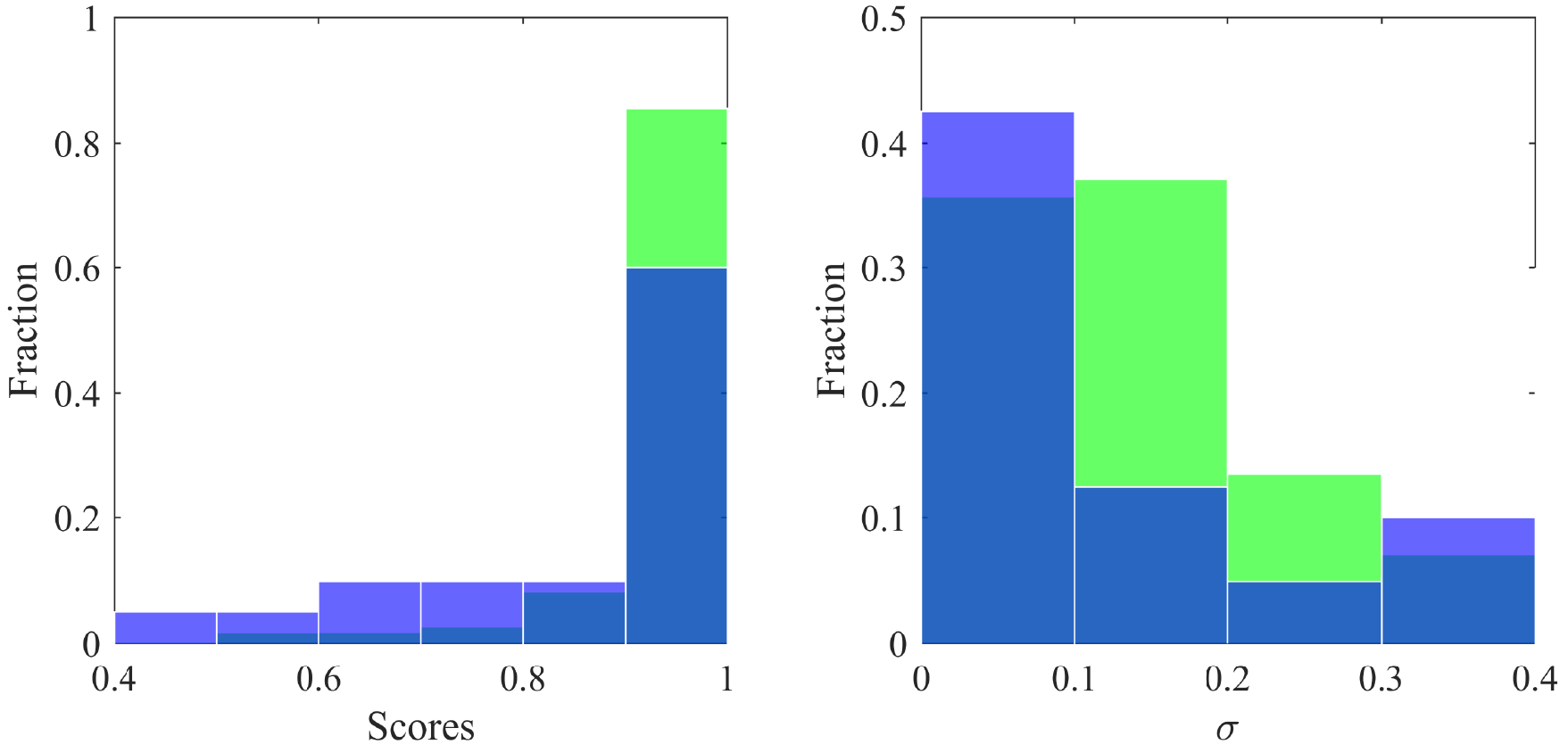}
   \caption{The classification result of the 6dFGS. Left panel: the distribution of the scores for correctly classified
            galaxies (green) and wrongly classified galaxies (blue). The one score means 100 percent.
            Right panel: the distribution of the scores' standard deviation.
   \label{SDFsc}}
\end{figure}

About 99.5\% of the galaxies are classified correctly, and 40 galaxies are wrongly classified as stars. Here all
the predicted QSOs are treated as the galaxies, since there is no QSO subtype in the 6dFGS. The classification
result is shown in Figure~\ref{SDFsc}. The scores of the correctly classified galaxies are larger than
those of the wrongly classified.
About 73\% of the correctly classified galaxies have $\sigma <$ 0.2, while only 55\% for the wrongly
classified galaxies have $\sigma <$ 0.2.
It indicates that the classifier is very uncertain about the types of the wrongly classified galaxies.

\subsubsection{RAVE}
The RAdial Velocity Extension (RAVE) is designed to provide stellar parameters to complement missions that focus on obtaining
radial velocities to study the motions of stars in the Milky Way¡¯s thin and thick disk and stellar halo \citep{Steinmetz06}.
Its fifth data release (DR5) contains 457,588 stars in the south sky \citep{Kunder17}.
There are 935 stars also observed by SDSS and $WISE$. We remove the stars used to build the classifier, and there are 737 stars left.
We feed the classifier with the nine colors and it yields 736 stars and one QSO with the accuracy of 99.9\%.
The wrongly classified star, SDSS J154142.28-194513.1, is located in the bright halo of the kap Lib. Its colors probably be
polluted by the bright star.

\begin{figure}
   \centering
   \includegraphics[width=0.45\textwidth]{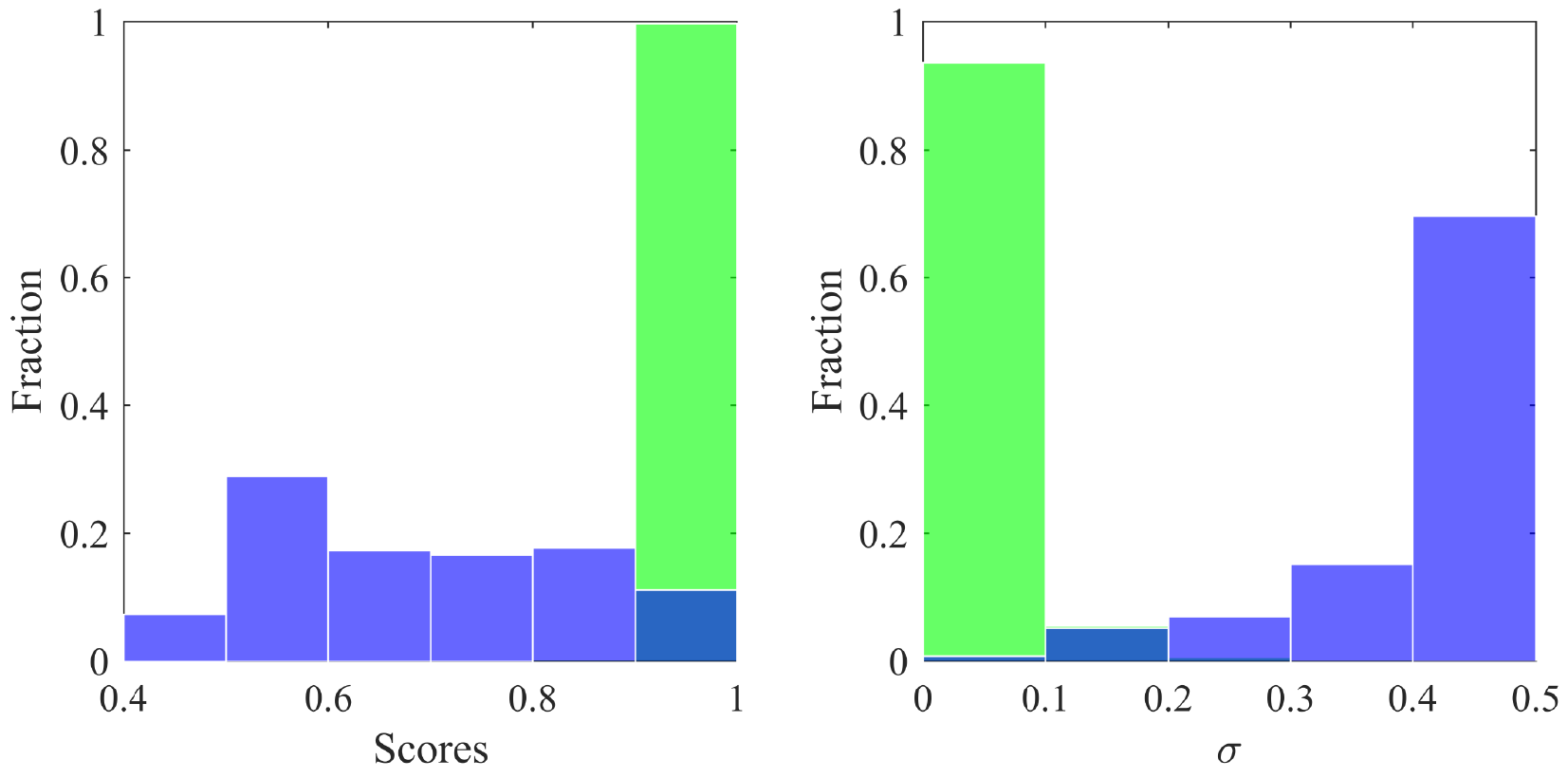}
   \caption{The classification result of the APASS-RAVE. Left panel: the distribution of the scores for correctly classified
            stars (red) and wrongly classified stars (blue).
            Right panel: the distribution of the scores' standard deviation.
   \label{RAVEsc}}
\end{figure}

We also take advantage the $g$, $r$ and $i$ magnitudes from APASS \citep{Munari14} that has been matched with RAVE stars \citep{Kunder17}.
Not all the stars are detected in both $WISE$ and APASS, and there are 435,012 stars with valid seven colors.
The prediction contains 434,735 stars, 264 galaxies and 13 QSOs with the accuracy of accuracy 99.9\%.
The classification result is shown in Figure~\ref{RAVEsc}.
The wrongly classified Stars have smaller scores and larger $\sigma$, implying the high uncertainty of the types.

\subsubsection{UVQS}
\begin{figure}
   \centering
   \includegraphics[width=0.45\textwidth]{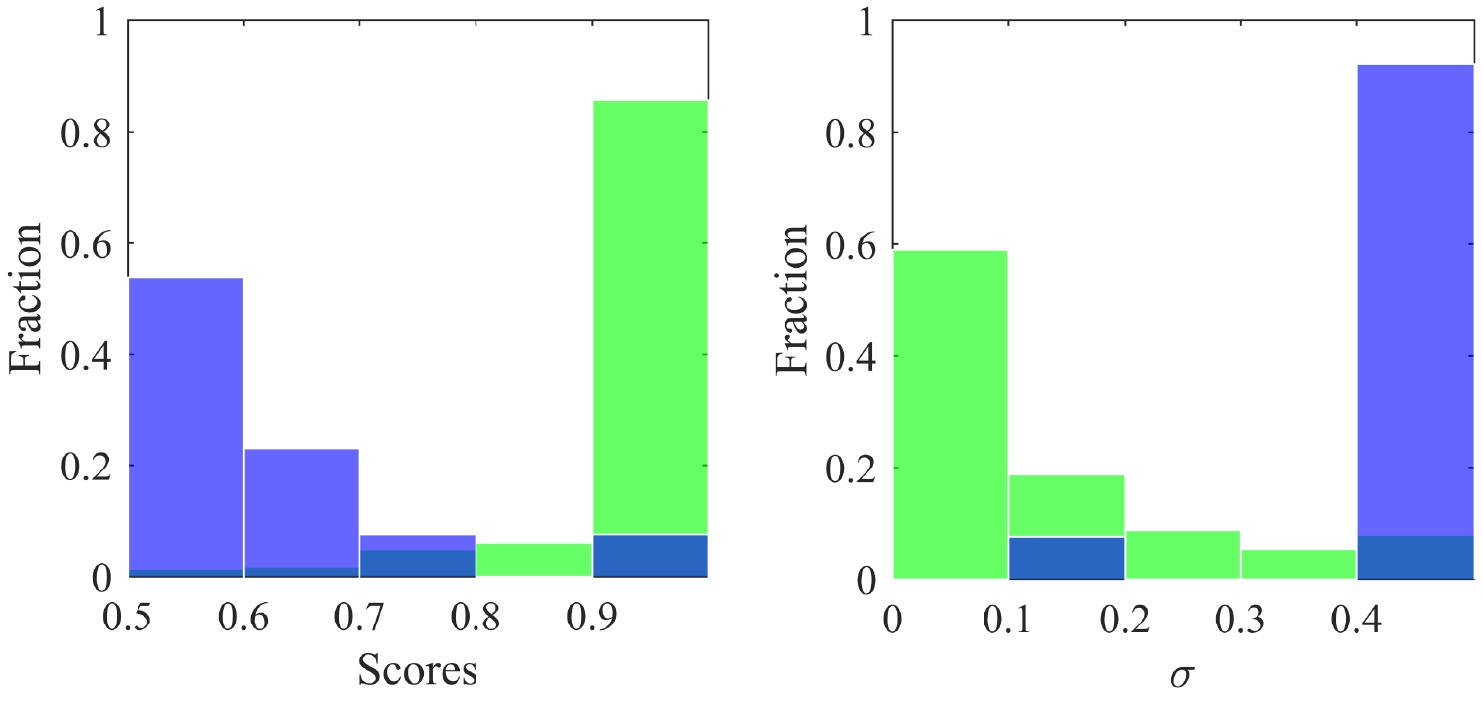}
   \caption{The classification result of the UVQS. Left panel: the distribution of the scores for correctly classified
            QSO (green) and wrongly classified QSOs (blue). Right panel: the distribution of the scores' standard deviation.
   \label{UVsc}}
\end{figure}

The data release one of all-sky UV-bright Quasar Survey (UVQS) contains 1,055 QSOs selected from $GALEX$ and $WISE$ photometry
and identified with optical spectra \citep{Monroe16}. We cross identified the QSOs with SDSS and $WISE$, which yields 262 QSOs.
We remove the QSOs used to build the classifier, and there are 237 QSOs left. The classifier yields 224 QSOs,
12 galaxies and one star with the accuracy of 94.5\%.
Again, the wrongly classified QSOs show smaller scores and larger $\sigma$ (Figure~\ref{UVsc}). The accuracies of the blind tests
is summarized in Table \ref{Table11}.

\begin{deluxetable}{cc}
\tablecaption{The accuracies of the blind test \label{Table11}}
\tablehead{Survey & Accuracy
           }
\startdata
6dFGS      & 99.5\%    \\
RAVE       & 99.9\%    \\
UVQS       & 94.5\%
\enddata
\end{deluxetable}

\section{Effective Temperature Regression}\label{regress}
We need more information on stars, after separating them from galaxies and QSOs.  The stellar spectral
classification organizes vast quantities of diverse stellar spectra into a manageable system, and has served as
the fundamental reference frame for the studies of stars for over 70 years \citep{Gray09}. In this section, we
present the method and the tests of our regression.

\subsection{Method}
\begin{figure}
   \centering
   \includegraphics[width=0.45\textwidth]{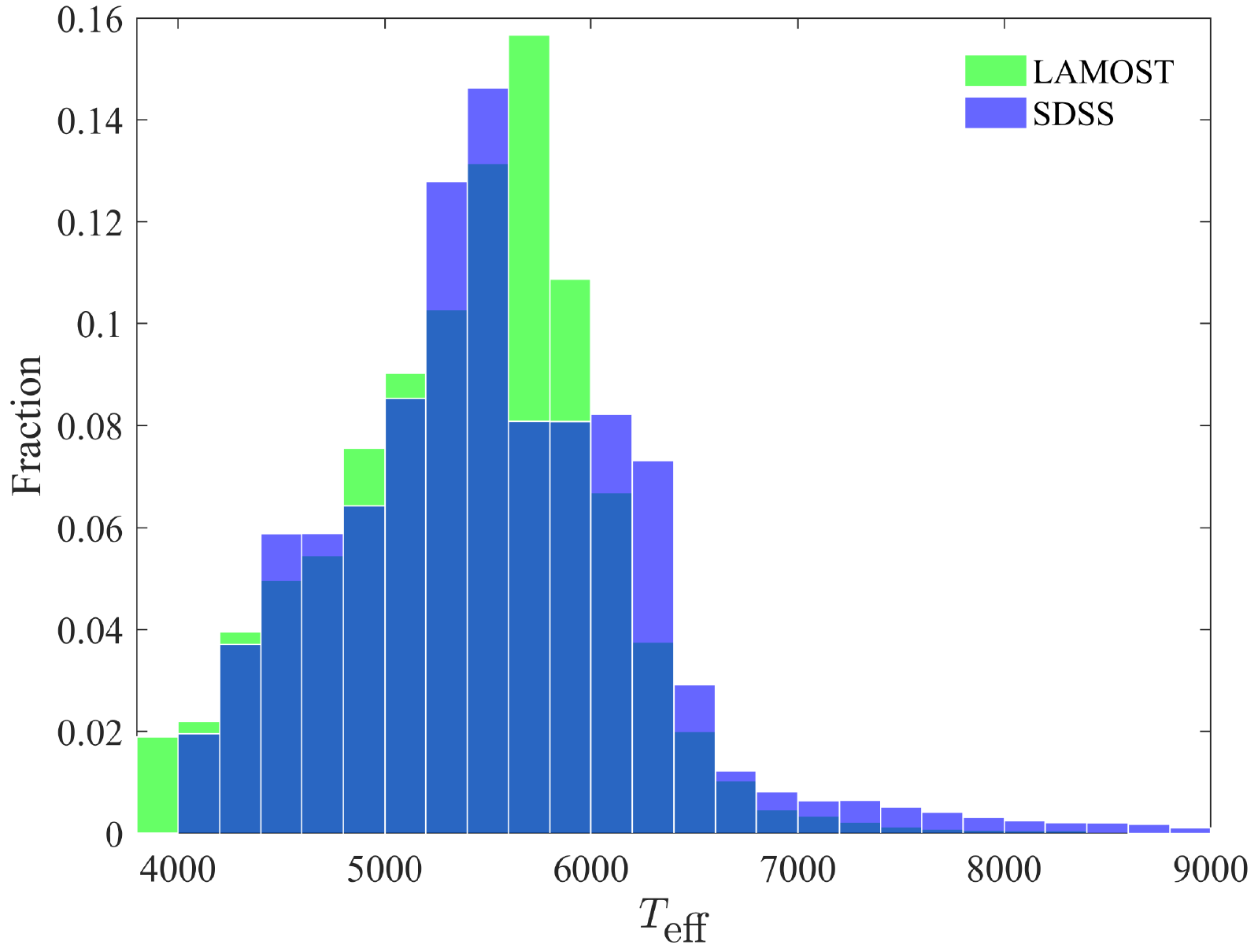}
   \caption{The normalized distribution of the {\Teff}. The green bars are the training sample of the LAMOST
            and the blue bars are that of the SDSS.
   \label{Teff}}
\end{figure}

The LAMOST¡¯s 1D pipeline only provides rough classification results and the accuracy of the subclasses is still not
robust \citep{Jiang13}.
Therefore, we instead adopt the effective temperatures (\Teff) from the A, F, G and K type star catalog, which was produced
by the LAMOST stellar parameter pipeline (LASP; \citealt{Wu14}).
We also extract the {\Teff} computed with the SEGUE Stellar Parameter Pipeline in the SDSS (SSPP;
\citealt{Allende08}; \citealt{Lee08a}; \citeyear{Lee08b}).
Both samples are dominated by G stars, and have similar distributions of {\Teff} (Figure \ref{Teff}).

In the classification (Secion \ref{class}), the RF exhibits advantage in the accuracy and the training time cost.
We here also adopt the algorithm of RF to build the regression of stellar effective temperature.

These temperatures and seven colors, $g-r$, $r-i$, $i-J$, $J-H$, $H-K$, $K-W$1 and $W$1$-$$W$2, of 1,327,071 stars
are used to train the RF for regression.
We apply the 10-fold cross validation in order to test the performance of the regression.
The cross validation partitions the sample into ten randomly chosen folds of roughly equal size.
One fold is used to validate the regression that is trained using the remaining folds. This process is repeated ten times such
that each fold is used exactly once for validation.

\begin{figure}
   \centering
   \includegraphics[width=0.45\textwidth]{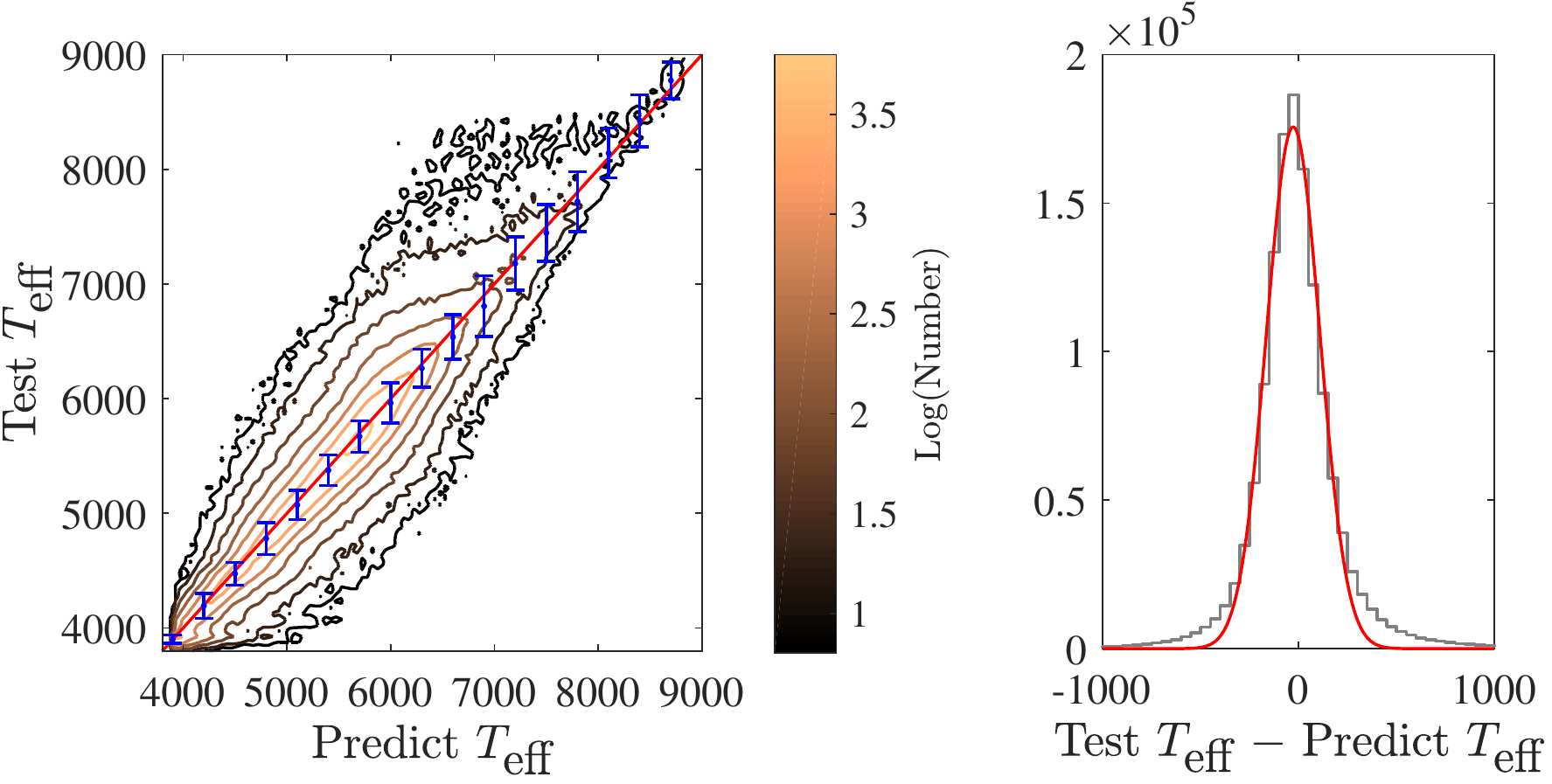}
   \caption{The one-one correlation of the regression (left panel). The blue error bars stand for the root-sum square
   of the standard deviation and the offset of the Gaussian fit in the bins of the predicted {\Teff}. The color bar stands for the
   color of the density contour in the log scale. The Gaussian fit (red) of the total residual (black) is shown in the right panel.
   \label{OOC}}
\end{figure}

We present the result of the cross validation in Figure \ref{OOC}. The one-to-one correlation is shown in the left
panel. In order to estimate the uncertainty of the prediction, we bin the predicted {\Teff} with a step size of 100 K,
and fit the distribution of the corresponding test {\Teff} with a Gaussian function. We calculate the root-sum square of
the standard deviation and the offset of the fit, which is adopted as the uncertainty of the prediction
(the blue error bars in Figure \ref{OOC}).
The Gaussian fit to the total residuals is shown in the right panel of Figure \ref{OOC}, and the fitted offset ($\mu$)
and the $\sigma$ are listed in Table \ref{Table2}. The red bars in Figure \ref{Imp} are the importance estimates for
the regression. The optical and 2MASS colors show much more importance than the $WISE$ colors, which are different from
those of the classification. It may be due to the majority of our sample that is G and K-type stars.

\subsection{Blind Test}
In this subsection, we use the {\Teff} extracted from the spectrum-based methods to test the actual performance of the regression.

\subsubsection{RAVE}
The RAVE pipeline processes the RAVE spectra and derives estimates of \Teff, log $g$, and [Fe/H] \citep{Kunder17}.
The pipeline is based on the combination of the MATrix Inversion for Spectral SynthEsis (MATISSE; \citealt{Recio06})
algorithm and the DEcision tree alGorithm for AStrophysics (DEGAS; \citealt{Bijaoui12}). This pipeline is valid for
stars with temperatures between 4000 K and 8000 K. The estimated errors in {\Teff} is approximately 250 K, and $\sim$ 100 K
for spectra with S/N \footnote{Signal-to-noise ratio in the RAVE database.} $\sim$ 50 \citep{Kunder17}.

\begin{figure}
   \centering
   \includegraphics[width=0.45\textwidth]{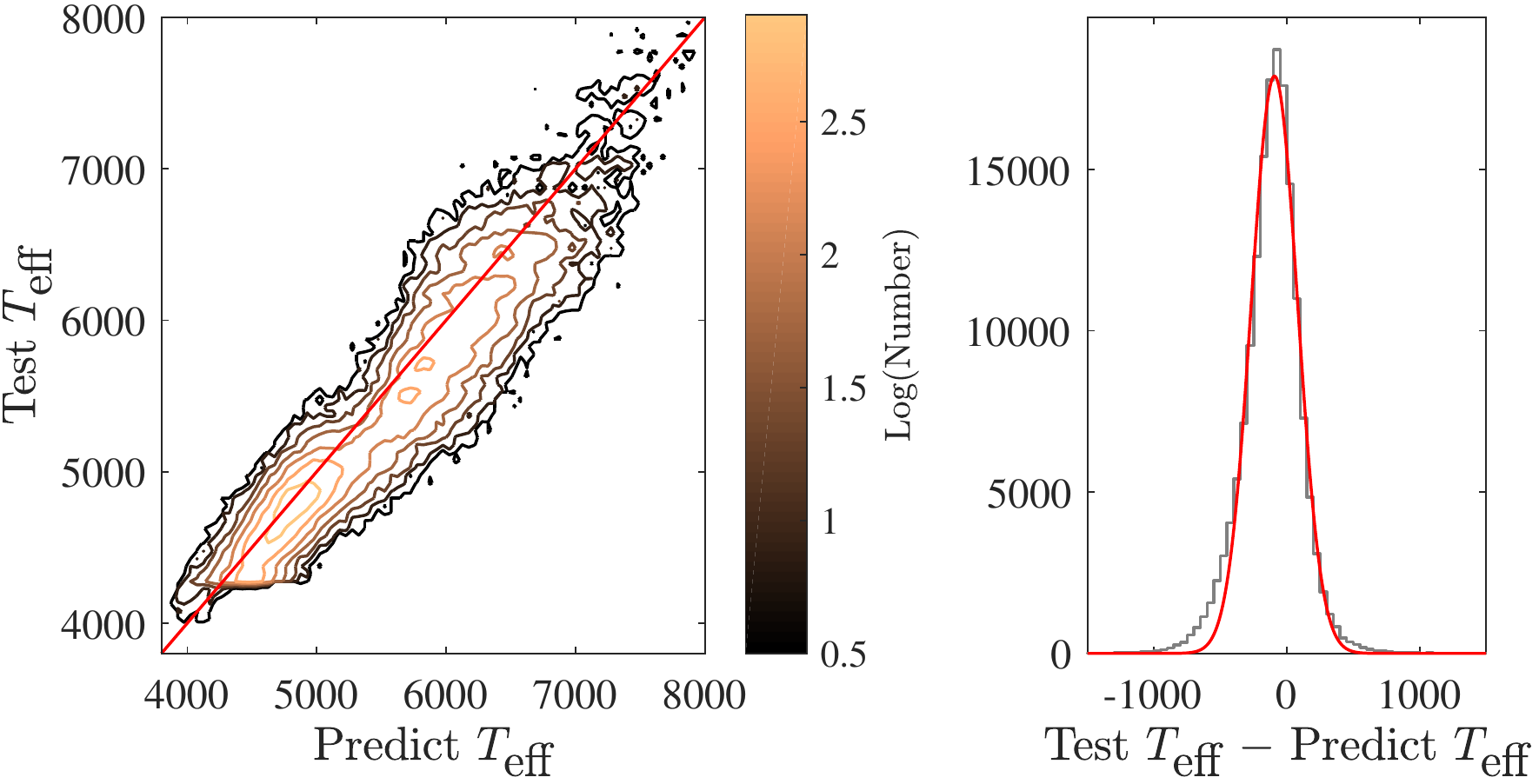}
   \caption{The one-one correlation of the stars in the APASS-RAVE (left panel).
            The Gaussian fit (red) of the total residual (black) is shown in the right panel.
   \label{OOCRAVE}}
\end{figure}

We adopt the photometry from APASS and $WISE$ in the RAVE database to construct the input colors.  The sample is restricted
to have S/N $\geqslant$ 50 and the quality flag of $Algo\_Conv$ $\neq$ 3 or 4 \footnote{The quality flag in the RAVE catalog, see Section 6.1 in \citet{Kunder17} for detail.}.
There are 165,011 stars left.
We present the prediction result in Figure \ref{OOCRAVE}, and list the parameters of the Gaussian fit to the total residuals
in Table \ref{Table2}.

\subsubsection{APOGEE}
The Apache Point Observatory Galactic Evolution Experiment (APOGEE), one of the programs in SDSS-III, has collected
high-resolution ($R$ $\sim$ 22,500) high signal-to-noise ($>$ 100) near-infrared (1.51$-$1.71 $\mu$m) spectra of
146,000 stars across the Milky Way \citep{Majewski17}. These stars are dominated by red giants selected from the
2MASS. Their stellar parameters and chemical abundances are estimated by the APOGEE Stellar Parameters and Chemical
Abundances Pipeline (ASPCAP; \citealt{Garcia16}). The typical error in {\Teff} is $\sim$ 100 K \citep{Meszaros13}.

\begin{figure}
   \centering
   \includegraphics[width=0.45\textwidth]{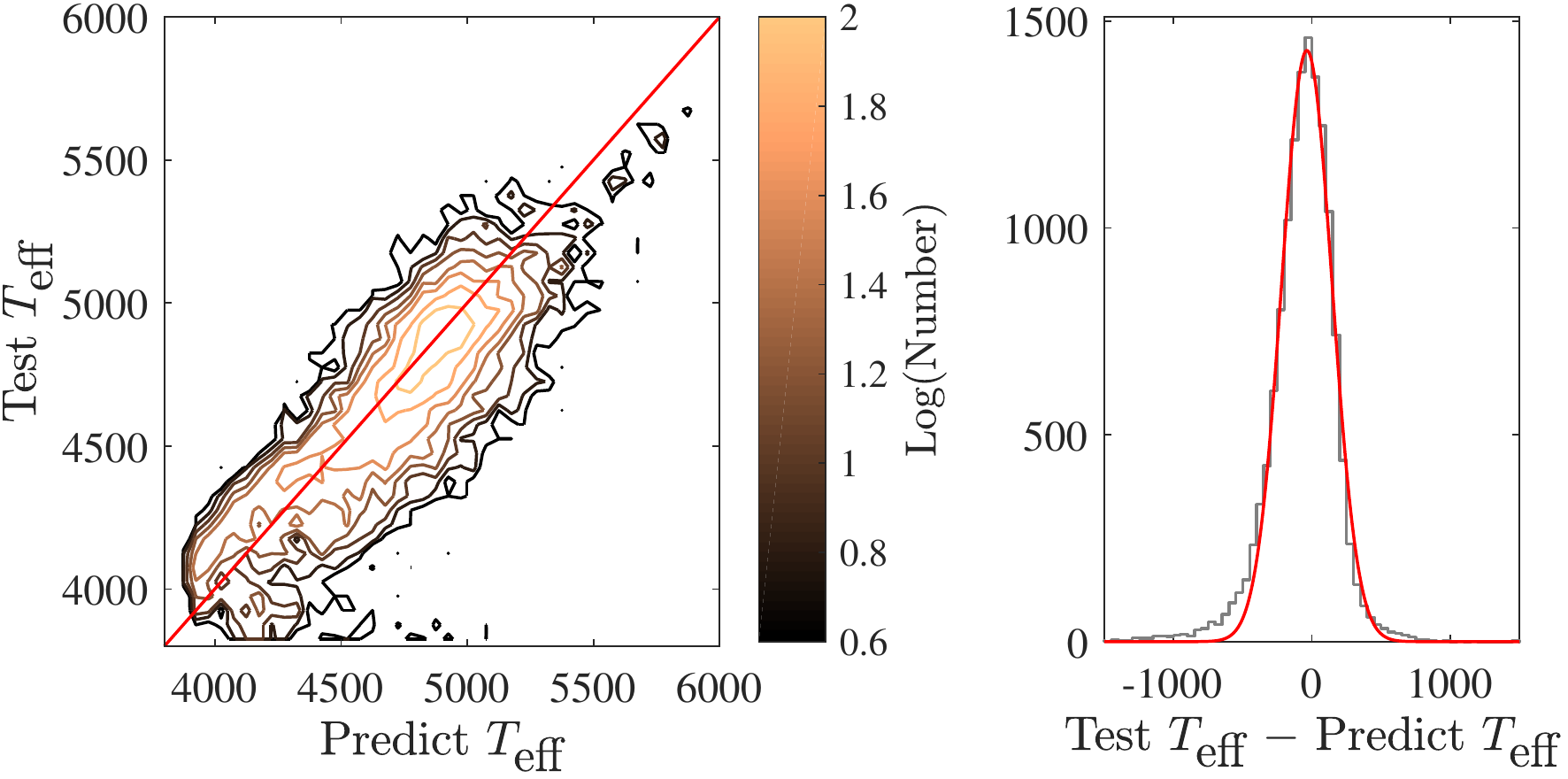}
   \caption{The one-one correlation of the stars in the APOGEE (left panel).
            The Gaussian fit (red) of the total residual (black) is shown in the right panel.
   \label{OOCAPO}}
\end{figure}

We extract the photometric data of SDSS and $WISE$ with the help of the Casjob. We feed the RF regression with the seven colors
of 13,685 stars. 
The prediction is shown in Figure \ref{OOCAPO}, and the parameters of fit to the total residuals are listed in Table \ref{Table2}.

We find that the offsets of the validation and the predictions are less than 100 K, and the standard deviations are
less than 200 K (Figure \ref{OOCRAVE} and \ref{OOCAPO}). \citet{Lee15} has applied the SSPP to LAMOST stars and compared the results to those from RAVE and APOGEE catalogs.
The offsets of {\Teff} between different pipelines are from 36 to 73 K and the standard deviations are from 79 to 172 K.
This indicates that our RF regression can determine the stellar temperatures with fair accuracy.

\begin{deluxetable}{l|cc}
\tablecaption{The Gaussian fits to the total residuals. \label{Table2}}
\tablehead{& $\mu$ & $\sigma$  \\
           & (K)   & (K)
           }
\startdata
Cross Validation       & $-$27 $\pm$ 2     & 136 $\pm$ 2 \\
RAVE                   & $-$93 $\pm$ 3     & 175 $\pm$ 3 \\
APOGEE                 & $-$36 $\pm$ 2     & 182 $\pm$ 2
\enddata
\end{deluxetable}

\section{Discussion}\label{dis}
The machine learning has been adopted as a successful alternative approach to defining reliable objects classes, stellar types
and types of variable stars (eg. \citealt{Liu15,Kovacs15,Krakowski16,Kuntzer16,Sarro18,Pashchenko18}).
It is not the first time to take advantage of this technology to classify the objects or to regress the stellar
parameters. In this section, we would like to compare our classification and regression to the results in other studies.

\subsection{Comparisons with other machine-learning methods}

\citet{Ball06} applied the supervised decision tree algorithm to classify the stars and galaxies in SDSS-DR3.
They used the colors $u-g$, $g-r$, $r-i$ and $i-z$ of 477,068 objects with spectroscopic attributes to train the
machine-learning classifier, and performed cross validation to test the performance.
The accuracy and completeness were over 90\%. Except for the optical colors, the IR colors are included in our multi-color data set,
since they have shown the importance in the machine-learning methods \citep{Henrion11}.
Our larger training sample and the IR aided color set result in a better performance of our classifier, over 99\%
for the stars and galaxies classification.
We also test the performance of some decision tree algorithms, and
the accuracies are $\sim$ 98\%. Compared to the decision tree, the random forest avoid overfitting to the training
set and limit the error from the bias \citep{Hastie08}.

\citet{Krakowski16} used the SVM learning algorithm to classify $WISE$ $\times$ SuperCOSMOS objects based on the SDSS
spectroscopic sources. The training sample included over one million objects, 95\% of which were galaxies, 2\% were stars,
and 3\% were QSOs. They used six parameters, $W$1, $W$3, $W$1$-$$W$2, $R-$$W$1, $B-R$ and w1mag13. The 10-fold cross
validation was performed to test the classifier, and the total accuracy was 97.3\%.
Instead of magnitudes, we adopt colors that are independent of distance. Our training sample shows better compositional balance,
and its size is three times larger than theirs.
We also try some SVM algorithms, and
the accuracies are from 70\% to 99\%. The time cost to build the SVM classifier is extremely longer than that for the RF classifier
\footnote{For example, the gaussian kernel SVM classifier has the highest accuracy among the SVM algorithms but cost over ten times longer than the RF classifier.}.
For a classification problem, the RF gives probability of belonging to classes \citep{Breiman01}, while the SVM relies on
the concept of "distance" between points that needs more time to calculate. The RF algorithm also shows better performance
than the SVM in other fields, such as \citet{Liu13}.

\citet{Liu15} employed an SVM-based classification algorithm to predict MK classes with 27 line indices measured from a small
sample, 3,134 LAMOST stellar spectra.
The holdout validation of 50\% was performed to test the accuracy of the classifier. The completeness
of A and G stars reached 90\%, while that of other stars was below 80\%. Since the spectral features of different types can
be very similar, clear cuts of these features probably lead to mis-classification. Therefore, we adopt the regression of {\Teff}
rather than the MK classification in order to avoid such effect.
\citet{Liu15}'s research also implies that a large sample could cover a larger area of the parameter space,
and further could yield more reliable prediction.


\citet{Sarro18} constructed regression models to predict {\Teff} of M stars with eight machine-learning algorithms.
The training sample is built with the features extracted from the BT-Settl of synthetic spectra. Then, the models were
applied to two sets of real spectra from the NASA Infrared Telescope Facility and Dwarf Archives collections.
\citet{Sarro18} used the root mean/median square errors (RMSE/RMDSE) to describe the prediction errors.
The RMSEs were from 160 to 390 K, and the RMDSEs from 90 to 220 K, various with the different algorithms
and signal-to-noise ratios.
Our prediction for A, F, G, K stars gives similar results: RMSE/RMDSE (RAVE) = 246/140 K, and RMSE/RMDSE (APOGEE) = 247/130 K,
implying that our regression built with photometric data could achieve similar accuracy to the spectrum-based model.

\subsection{SED Fit}
\begin{figure}
   \centering
   \includegraphics[width=0.5\textwidth]{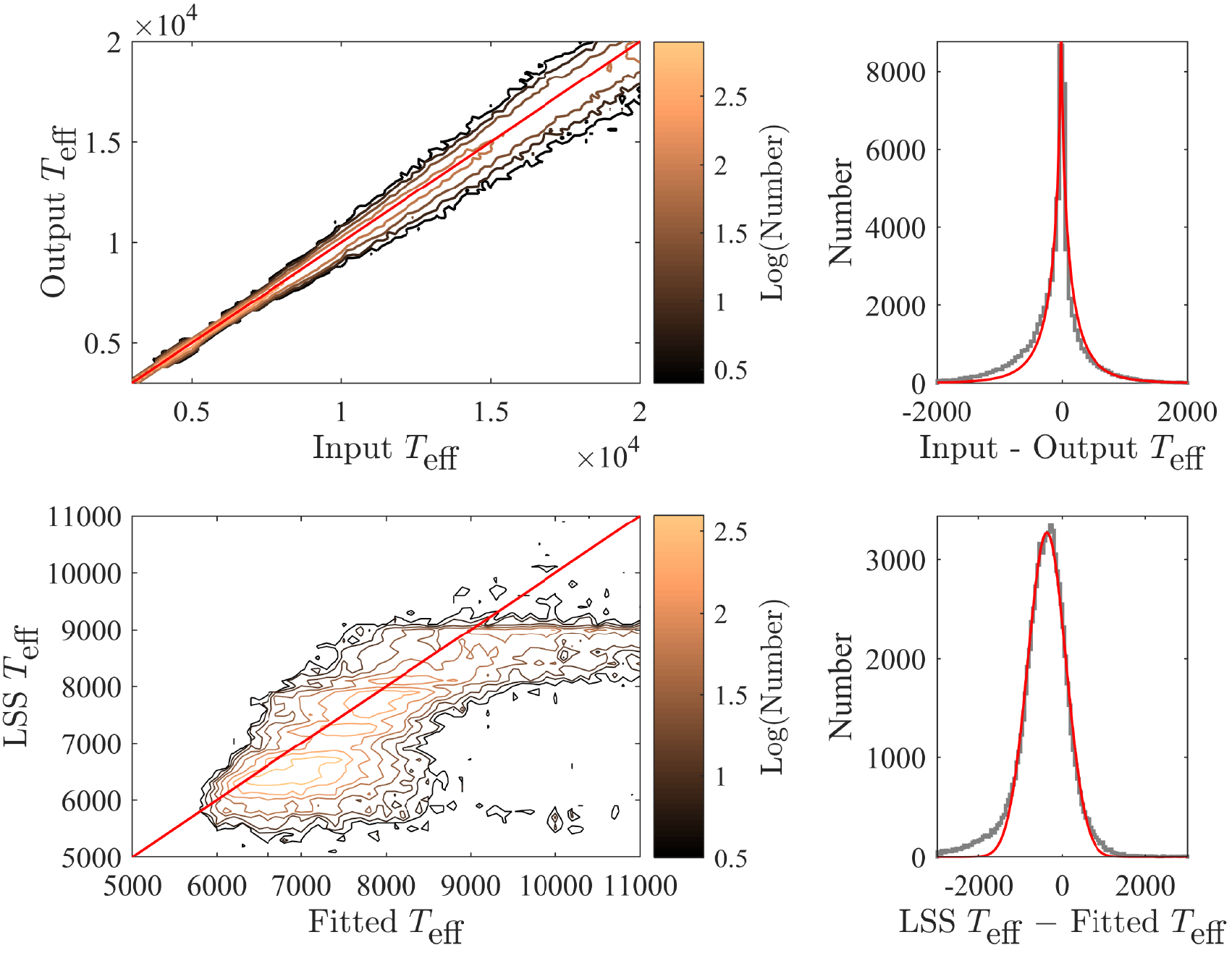}
   \caption{The results for the theoretical simulation (the upper panels) and the application to LSS-GAC catalog (the lower panels).
            The color bars stand for the colors of the density contours. We use the Gauss-like function (the red line) to fit the
            total residual of the theoretical simulation, and use the Gaussian function to fit the residual of the stars in the LSS-GAC catalog.
   \label{OI}}
\end{figure}

Another way to determine the {\Teff} is the fit of stellar SEDs with synthetic templates.
The theocratical study has concluded that the broad-band photometry from the UV to the mid-IR allows atmospheric parameters and
interstellar extinction to be determined with good accuracy \citep{Allende16}. The study used the SEDs extracted from the
ATLAS9 model atmospheres \citep{Meszaros12}. They added interstellar extinction to these SEDs in order to construct the theocratical templates.
The test SEDs were also extracted from the ATLAS9 model, but added some random noise.
Then the test SEDs were fitted with the templates using the $\chi^2$-optimization method. The standard deviations of the total residual
were from 130 to 380 K depending on different bands used for the fittings. We follow this procedure to fit 10$^5$ simulated SEDs,
extracted from the BT-Cond theoretical model \citep{Baraffe03,Barber06,Allard10}.
Since the simulated SEDs have random stellar parameters, about 70,000 SEDs are located inside the reasonable ranges.
The result is shown in the upper panels in Figure \ref{OI}.
We only plot $\sim$ 64,000 samples with $\chi^{2}$ $\leqslant$ 5.88 that is one standard deviation for a $\chi^{2}$ distribution with the five
degrees of freedom ({\Teff}, log $g$, [Fe/H], $E(B-V)$ and the scaling factor).
The residuals are fitted with a Gauss-like function:
\begin{equation}
f = Ae^{\frac{-(x-\mu)^\delta}{\delta\sigma^\delta}}.
\end{equation}
The standard deviation of {\Teff} is 207 $\pm$ 15 K for 12 bands fit, F-NUV, $ugriz, JHK, W1W2$.
The standard deviation of other parameters are also similar to the result in \citet{Allende16},
indicating that the multi-band SED fit can well constrain the atmospheric parameters and interstellar extinction theocratically.

However, the result is worse than expected (the lower panels in Figure \ref{OI}), when we apply the SED templates to
fit the SEDs in the real observation, the stars in the LAMOST Spectroscopic Survey of the Galactic Anticentre (LSS-GAC;
\citealt{Liu14}; \citealt{Yuan15}).  The standard deviation of {\Teff} is 454 $\pm$ 5 K and the offset is $-$365 $\pm$ 5 K,
larger than those of the machine-learning regression by a factor of three.
We also try this technology in other ways, e.g., fitting the stars in RAVE, or using 10 bands fit \footnote{Some study have shown that the UV emission is from the
higher regions of the stellar atmosphere and lead to discrepancies between observations and the theoretical models \citep{Bai18a}.}.
The standard deviations of {\Teff} are about 400 K, extremely worse than both the theoretical simulation and the machine-learning regression.
This implies that the atmospheric parameters of the stars in the real observation can't be well estimated by the SED fit using the $\chi^2$ minimization.
Based on photometric data, machine learning shows better performance on the {\Teff} estimate.

\subsection{A Scientific Application}
The ESA space mission $Gaia$ is performing an all-sky survey at optical wavelengths, and its primary objective is to survey
more than one billion stars \citep{Gaia16}. Its second data release ($Gaia$ DR2; \citealt{Gaia18}) includes $\sim$ 1.3 billion objects with valid
parallaxes. These parallaxes are obtained with a complex iterative procedure, involving various
assumptions \citep{Lindegren12}. Such procedure may produce parallaxes for galaxies and QSOs, which
should present no significant parallaxes \citep{Liao18}.

We have applied the classifier to 85,613,922 objects in the $Gaia$ DR2 based on the multi-wavelength data from Pan-STARRS and $WISE$ \citep{Bai18b}.
The result shows that the sample is dominated by stars, $\sim$ 98\%, and galaxies and QSOs make up 2\%. For the objects with negative parallaxes,
about 2.5\% are galaxies and QSOs. About 99.9\% of the sample are stars if the relative parallax uncertainties are smaller than 0.2,
implying that using the threshold of 0 $< \sigma_\pi/\pi <$ 0.2 could yield a very clean stellar sample \citep{Bai18b}.

\section{Summary and Future Work}\label{sum}
In this work we have attempted to classify the objects into stars, galaxies and QSOs, and further regress the effective temperatures
for stars using the machine learning, the algorithm of choice being the RF. The classifier is trained
with about three million objects in SDSS, LAMOST and Veron13, and the regression is trained with one million stars in SDSS and LAMOST.
In order to exam the performance of the classifier, we perform three blind tests by using objects spectroscopically identify
in the RAVE, 6dFGS, and UVQS. The total accuracies are over 99\% for the RAVE and 6dFGS, and higher than
94\% for the UVQS. We also perform two blind tests for the regression by using the stellar {\Teff} estimated with spectroscopical
pipelines in the RAVE and APOGEE. The offsets and the standard deviations of the total residual are below 100 K and 200 K, respectively.

Our classifier shows the high accuracy compared to other machine-learning algorithms in former studies, indicating that combining broad-band photometry
from the optical to the mid-infrared allows classification to be determined with very high accuracy. The machine learning provides us an efficient approach
to determine the classes for huge amounts of objects with photometric data, e.g., over four hundred million objects in the SDSS-$WISE$ matched catalog.

Since there is no clear cut for colors or spectral features of the different spectral types, we adopt {\Teff} regression rather
than the MK classification to further provide basic information on stars.
Our regression result shows similar or even better performance than the SED $\chi^2$ minimization and some spectrum-based methods.
The RF regression enable us to estimate the {\Teff} without spectral data for the stars that are too many or too faint for the spectral
observation, or the stars in the large area time dominated survey (e.g., Pan-STARRS1 survey; \citealt{Chambers16}).

We are going to test regressions for other stellar parameters with machine-learning algorithms.
We also try to decouple the effective temperature and the
interstellar extinction based on large sample, such as LAMOST-SDSS-Gaia.
The future well controlled sample, e.g., LAMOST-II and SDSS-V \citep{Kollmeier17}, also provides us an opportunity to explore
the multi-dimensional parameter space with this technology for classification and regression.

\vspace{3ex}
The machine-learning results in this work are developed with MATLAB\footnote{https://www.mathworks.com} available upon request to the first author as MAT files.

\begin{acknowledgements}

We are grateful to Stephen Justham for valuable discussions.
This work was supported by the National Program on Key Research and Development
Project (Grant No. 2016YFA0400804) and
the National Natural Science Foundation of China (NSFC)
through grants NSFC-11603038/11333004/11425313/11403056.
Some of the data presented in this paper were obtained from the Mikulski Archive for
Space Telescopes (MAST). STScI is operated by the Association of Universities for Research
in Astronomy, Inc., under NASA contract NAS5-26555. Support for MAST for non-HST data is provided
by the NASA Office of Space Science via grant NNX09AF08G and by other grants and contracts.

The Guoshoujing Telescope (the Large Sky Area Multi-Object
Fiber Spectroscopic Telescope, LAMOST) is a National Major
Scientific Project which is built by the Chinese Academy of
Sciences, funded by the National Development and Reform Commission,
and operated and managed by the National Astronomical Observatories,
Chinese Academy of Sciences.

Funding for the Sloan Digital Sky Survey IV has been provided by the Alfred P.
Sloan Foundation, the U.S. Department of Energy Office of Science, and the
Participating Institutions. SDSS-IV acknowledges
support and resources from the Center for High-Performance Computing at
the University of Utah. The SDSS web site is \url{http://www.sdss.org/}.

SDSS-IV is managed by the Astrophysical Research Consortium for the
Participating Institutions of the SDSS Collaboration including the
Brazilian Participation Group, the Carnegie Institution for Science,
Carnegie Mellon University, the Chilean Participation Group, the French
Participation Group, Harvard-Smithsonian Center for Astrophysics,
Instituto de Astrof\'isica de Canarias, The Johns Hopkins University,
Kavli Institute for the Physics and Mathematics of the Universe (IPMU) /
University of Tokyo, Lawrence Berkeley National Laboratory,
Leibniz Institut f\"ur Astrophysik Potsdam (AIP),
Max-Planck-Institut f\"ur Astronomie (MPIA Heidelberg),
Max-Planck-Institut f\"ur Astrophysik (MPA Garching),
Max-Planck-Institut f\"ur Extraterrestrische Physik (MPE),
National Astronomical Observatories of China, New Mexico State University,
New York University, University of Notre Dame,
Observat\'ario Nacional / MCTI, The Ohio State University,
Pennsylvania State University, Shanghai Astronomical Observatory,
United Kingdom Participation Group,
Universidad Nacional Aut\'onoma de M\'exico, University of Arizona,
University of Colorado Boulder, University of Oxford, University of Portsmouth,
University of Utah, University of Virginia, University of Washington, University of Wisconsin,
Vanderbilt University, and Yale University.
\end{acknowledgements}

\end{document}